\documentclass[12pt, letterpaper]{article}
\usepackage{setspace}
\usepackage{geometry}
\usepackage{titlesec}
\usepackage{fancyhdr}
\usepackage{lipsum}
\usepackage{datetime}
\usepackage{titling}
\usepackage{authblk}
\usepackage{graphicx}

\renewcommand{\maketitle}{
  \begin{titlepage}
    \begin{center}
      \vspace*{.8cm}
      \LARGE{\textbf{Is the Page-time paradox paradoxical?}}\\
      \vspace{.8cm}
      \Large{Bruno Arderucio Costa\textsuperscript{*}}\\
      \texttt{arderucio@alumni.ubc.ca}\\
      \vspace{.8cm}
      \large{Instituto de Ciencias Nucleares, Universidad Nacional Autonoma de Mexico\\
 Apartado Postal 70-543, Cd.~Mx., 04510, Mexico;\\
 Center for Relativity and Cosmology at Troy University, AL, USA}\\
      \vspace{.8cm}
      \today\\
      \vspace{1cm}
      \begin{minipage}{0.8\textwidth}
        \small{
          \textbf{Abstract:} I discuss how five reasonably sounding assumptions lead to a dilemma---the Page-time paradox---, which appears to challenge a conventional statistical mechanical underpinning of black hole thermodynamics. By inspecting the conceptual subtleties behind each hypothesis, I list questions that require clarification before the puzzle can be deemed paradoxical. I devote particular attention to using thermodynamic arguments for a system that never reaches equilibrium. As a proof of concept, I show that the paradox is absent in a modified setting that admits an equilibrium thermodynamics formulation. 
          \vspace{0.5cm}
          
          Essay written for the Gravity Research Foundation 2024 Awards for Essays on Gravitation.
        }
      \end{minipage}
      \vfill
      \small{\textsuperscript{*} Corresponding author}
    \end{center}
  \end{titlepage}
}

\titleformat{\section}{\normalfont\Large\bfseries}{}{0em}{}

\begin{document}
\title{}
\maketitle

Charles Misner attributes the following quote to John Wheeler~\cite{Misner06}:
\begin{quote}
    An expert is one who has made every mistake in some field.
\end{quote}

Indeed, it is difficult to overstate how effective learning from one's mistakes is, so much so that learning from other's mistakes is a valuable shortcut. The apparent paradoxes in special relativity, such as the twin and pole--barn paradoxes, epitomise mistake catalysts. Even though one can, on general grounds, know that there are no inconsistencies in special relativity, comprehending exactly where the reasoning that led to an illusive paradox failed adds a new layer of depth to our understanding. Such a layer is invaluable for honing our intuitions about, in the case of the pole--barn paradox, the relativity of simultaneity. The various formulations of the so-called `information paradox' in black hole physics serve similar purposes. In its most naive form, the paradox is an alleged contradiction between the destruction of information by the singularity of a black hole that fully evaporates and the unitary time evolution of quantum mechanics. This formulation reminds us of the importance of the well-posedness of an initial value formulation.

\section{The conundrum}
This essay focuses on a related, more sophisticated paradox arising from black hole evaporation, reviewed in Ref.~\cite{almheiri21}. To distinguish it from the most famous version, I adopt the terminology introduced by Wallace~\cite{wallace18}, the \emph{Page-time paradox}. In rough terms, the Page-time paradox reveals that the assumption that the thermodynamic entropy of an evaporating black hole can be accounted for by the entanglement entropy of the matter across the horizon is self-contradictory. I formulate the paradox as follows:

The following five propositions are mutually incompatible:
\begin{enumerate}
    \item The von Neumann entropy of the reduced state of the quantum matter localised in the interior of a black hole, $S_\mathrm{int}$, is always equal to the von Neumann entropy of the reduced state of its exterior, $S_\mathrm{ext}$.
    \item The thermodynamic entropy (i.e., the entropy that enters in the first and second laws of thermodynamics) of a black hole, $S_\mathrm{th}$, satisfies $S_\mathrm{th}\geq S_\mathrm{int}$.
    \item (a) When matter is in a suitable quantum state, a black hole can lose a substantial fraction of its mass through Hawking radiation. In such a state, the emission of Hawking radiation is approximated by that of a black body in the same background geometry describing the black hole. (b) The mass of the matter in the exterior increases to preserve global energy conservation.
    \item The thermodynamic entropy of a black hole is \emph{always} equal to, or smaller than, the Bekenstein--Hawking entropy $S_\mathrm{BH}$\footnote{In Planck units, $S_\mathrm{BH}$ is a quarter of the area of the event horizon intercepted with a Cauchy surface, which I assume to exist in this essay since the contradiction is present even if the black hole does not evaporate completely. Even though the Bekenstein--Hawking formula is rooted in general relativity, this proposition could be equally applied to other theories of gravity as long as their corresponding entropy decreases as the black hole evaporates.}, i.e. $S_\mathrm{th}\leq S_\mathrm{BH}$.
    \item The entropy $S_\mathrm{ext}$ of the matter surrounding a black hole is well defined and a monotonically increasing function of its energy.
\end{enumerate}

I motivate the individual premises below, but first, I draw out the inconsistency. Propositions 1 and 2 yield $S_\mathrm{ext}=S_\mathrm{int}\leq S_\mathrm{th}$. From propositions 3(a) and 4, we conclude that $S_\mathrm{th}$ reduces to arbitrarily small values as the black hole evaporates. Thus, $S_\mathrm{ext}$ should also become arbitrarily small. However, according to 3(b) and 5, $S_\mathrm{ext}$ should increase as the black hole evaporates and eventually become larger than $S_\mathrm{BH}$. The time coordinate at which the equality $S_\mathrm{ext}=S_\mathrm{BH}$ is reached is called the Page time. Beyond the Page time, the five propositions above can be used to establish that $S_\mathrm{ext}$ either grows larger than $S_\mathrm{BH}$ or is forever bounded from above by $S_\mathrm{BH}$, a clear contradiction.

The contradiction reveals that at least one of the statements above is false. Whichever that is, its identification will unequivocally teach us something unexpected.

\section{Premises' Strengths and Limitations}
The first proposition arises naturally if one assumes that the initial state, before the formation of a black hole, is pure\footnote{It is easy to build an intuition about why by studying a bipartite system described by a finite-dimensional Hilbert space $\mathcal H=\mathcal H_A\otimes\mathcal H_B$. Assuming that the composite state is pure, one uses the Schmidt decomposition to write the state vector as a linear combination of a product basis and ascertain that the von Neumann entropies associated with reduced states in $L(\mathcal H_A)$ or $L(\mathcal H_B)$ are equal. The situation is more complicated when dealing with quantum fields because the entropy of either reduced state is infinite. Yet, it is unlikely that this poses an insurmountable problem since the origin of this divergence is well understood~\cite{Longo21}.

Furthermore, there is mounting evidence~\cite{witten22} that, once gravity is included, unlike in quantum field theory in a fixed background, the entropy may be finite.}. Naturally, one could start with a mixed state and deny the first premise. However, with no independent reason to forbid pure states, it would be shocking if the Page-time paradox declared them unphysical.

The second proposition is commonplace in statistical mechanics. One can define the thermodynamic entropy as the maximum value of an entropy functional subjected to certain constraints and derive the laws of thermodynamics from them~\cite{caticha08}. In this framework, the thermodynamic entropy is, by definition, the value of the entropy when the system is in equilibrium. The procedure can be justified on the precept that it is undesirable to embrace an assumption in the absence of evidence to support it. The entropy functional is a measure of ignorance and, as such, should be maximised to infer the most honest probability distribution. Hence, when the entropy is evaluated at any other state compatible with the given constraints, its value cannot exceed the thermodynamic entropy.

Yet, two subtleties cast doubt on the present application. First, we identify the thermodynamic entropy when there is a conditional maximum for the entropy functional, which identifies with the equilibrium state. However, the interior of a black hole is not stationary, meaning that matter can never be expected to equilibrate. Thus, it is unclear whether the entropy functional has the required maximum. Second, the chronological past of the inextendible worldlines representing any observer outside the black hole never intercepts its interior. Insofar as the exterior, where the laws of black hole thermodynamics apply, is concerned, no interior information is available to serve as constraints (except for the standard normalisation constraint). 

A not-so-subtle objection is that while it is true that the maximisation procedure leads to a thermodynamic entropy, it is not true that \emph{any} thermodynamic entropy results from this procedure. For example, the Minkowski vacuum for a scalar field (with an ultraviolet cutoff) is a thermal state of zero temperature. This is an extensive system whose ground state is non-degenerate. Therefore, we conclude from Nernst's theorem~\cite{Landau5} that the thermodynamic entropy of any of its parts is zero. However, the entanglement entropy between the interior and exterior of an imaginary sphere in Minkowski spacetime is positive~\cite{Bombelli86, Srednicki93}, $S_\mathrm{int}>S_\mathrm{th}=0$. This example evinces that black holes' adherence to the laws of thermodynamics does not, by itself, support premise 2.

Even though the example above is not a black hole, violations of premise 2 are present in any system whose key features are (i) non-degeneracy of the ground state, (ii) positive heat capacity, which is automatic for extensive systems in the thermodynamic limit, and (iii) the presence of entanglement between the two parts into which a pure state is subdivided. For these reasons, I regard premise 2 as the least substantiated among the five, and one can argue that the contradiction counts as evidence that it should be rejected. There are, nonetheless, other means to avoid a paradox, which I explore below.

The third claim is not free of technical challenges. For example, the existence of a state that satisfies (a) and (b) is not generally guaranteed in quantum field theory in a fixed background, and even so, nobody knows how gravity responds to quantum matter. In classical general relativity, it is a consequence of Einstein's equations that the Arnowitt--Deser--Misner (ADM) mass, a measure of the `total' mass of asymptotically flat spacetimes, is conserved. For stationary spacetimes, one can decompose the ADM mass in terms that depend on the energy-momentum tensor on the exterior of the black hole and an integral over the horizon. It is then natural to identify how much a black hole contributes to the ADM mass. Thanks to Einstein's equations, we rest assured that if the black hole shrinks, the mass in its exterior must compensate for that. When matter demands a quantum description, one possibility is that, at least approximately, the geometry is sensitive only to the expectation value of the energy-momentum tensor after renormalisation (semiclassical gravity), which would mimic the energy conservation in classical theory. It would be astonishing if a gravity-matter coupling more accurate than semiclassical gravity were such that no reminiscent of the ADM mass conservation remained.

Even though assumption 3 is not rigorously stated, it encapsulates our physical expectations and could be used to assess how reasonable a contending mathematical theory is. As interesting as these technical obstacles are, premise 3(b) is likely out of danger. And so is 3(a), which receives support from quantum field theory fixed backgrounds~\cite{Hawking75, Unruh76}.

The fourth, innocent-looking premise is subtle. General relativity in stationary spacetimes permits the identification of the Bekenstein--Hawking entropy as a black hole's thermodynamic entropy. However, the situation is more delicate for an evaporating black hole. Recall that the laws of standard thermodynamics are formulated for systems that transition from one equilibrium state to another. When an archetypal textbook application of the first law does not downright assume a quasi-static transformation, one can employ a surrogate transformation that shares the initial and final states with the physical process to compute the difference in thermodynamic potentials that are a function of the state.

The laws of black hole thermodynamics are no different. There are two versions of the first law. First, a `local', or sometimes `physical process', version~\cite{Hawking72} computes the fluxes of energy and angular momentum across the horizon and relates them with the change in the area. Critically, it assumes that the final state is stationary. Second, a `global' version that compares parameters describing a family of stationary, asymptotically flat spacetimes~\cite{bch}. This `global' version is a closer analogue of ordinary thermodynamics and serves as a consistency check for the local version. The generalised second law of thermodynamics cannot be formulated in stationary spacetimes because it is a statement about how entropy changes in time. It, too, admits a `local' and a `global' version. The former applies to dynamical situations~\cite{wall12} but is insufficient to determine the value of the entropy uniquely~\cite{Hollands24}. To establish the global version in semiclassical gravity, one must be able to define the initial and final entropies. This is possible if one assumes that the spacetime transitions from one stationary epoch to another~\cite{bac20}\footnote{One does not need to adopt the proposed dynamical definition for the entropy in that reference. It suffices to agree that whichever definition for the dynamical entropy one adopts should coincide with the usual formula in the stationary epochs (after a certain `relaxation time').}. The necessity of such an assumption should come as no surprise because the same is true in ordinary thermodynamics! To prove that the entropy grows \emph{in time}, one must assume that the final state is an equilibrium state~\cite{jaynes65}. 

While there is no shortage of candidates for a statistical mechanical description of black holes~\cite{Carlip_2009}, the Bekenstein--Hawking formula is grounded in the first law of black hole mechanics and the generalised second law of thermodynamics, much like usual thermodynamics before Boltzmann and Gibbs. Only by leaning on a thermodynamic equilibrium can we confidently identify variations in entropy (see Ref.~\cite{prunkl19} for a discussion). However, black hole evaporation in premise 3(a) conditions does not meet this criterion and deserves closer inspection (next section).

Yet, in defence of premise 4, if a definition of the non-stationary thermodynamic entropy differs appreciably from the Bekenstein--Hawking formula in general relativity, we expect, in the spirit of statistical mechanics, its value to be smaller than $S_\mathrm{BH}$, which does not help unravel the riddle.

The fifth hypothesis merely states that nothing out of the ordinary is expected from the behaviour of the matter that originated from Hawking radiation. Yet, abandoning this premise is how the authors of Ref.~\cite{almheiri21} propose to solve the puzzle. I do not consider premise 5 pivotal in the Page-time paradox. Indeed, one could replace it with
\begin{enumerate}
    \item[5'.] The generalised entropy $S_\mathrm{gen}=S_\mathrm{th}+S_\mathrm{ext}$ cannot decrease in time, i.e., the value of $S_\mathrm{gen}$ on an arbitrary Cauchy surface $\Sigma_2$ will be at least the value of $S_\mathrm{gen}$ on a Cauchy surface $\Sigma_1$ in the past of $\Sigma_2$. 
\end{enumerate}
and reach a similar contradiction. From premises 1 and 2, $S_\mathrm{gen}\leq 2S_\mathrm{th}$. However, from premise 4, $S_\mathrm{th}$ becomes arbitrarily small for the state described by premise 3(a), which would entail a breakdown of 5'. Incidentally, in this formulation of the paradox, assumption 3(b) is expendable.

\section{Forced Equilibration}
As argued before, the paradox may stem from the limitations in the applicability of thermodynamics to a system that never thermalises. A test for this hypothesis is to force the system back into equilibrium. One can achieve this with black holes by confining them within sufficiently small boxes~\cite{Hawking76}. We begin with a large black hole and no matter inside the box. If, when the system reaches equilibrium, the entropy of the matter exceeds the value of the entropy of the black hole, the Page-time paradox would prevail over our setting modification, suggesting that we should turn our attention to other possible resolutions for the paradox. We perform an analysis similar to Hawking's~\cite{Hawking76} as follows: suppose the matter obeys an equation of state of the form
\begin{equation}
    E_m=\alpha V T^r,
    \label{eostate}
\end{equation}
where $E_m$ denotes the total energy of the matter, $V$ its volume, $T$ the temperature, and $\alpha>0$ and $r$ are constants. The total entropy of the combined system is $S=S_m(E_m)+S_{bh}(E_{bh})$, that is, the sum between the entropy $S_m$ of the matter in the exterior, which is a function of $E_m$ alone, and the entropy $S_{bh}$ of the black hole, which is a function of its energy $E_{bh}$. For simplicity, we take a Schwarzschild black hole, for which $S_{bh}(E_{bh})=\frac{1}{4}4\pi(2E_{bh})^2=4\pi E_{bh}^2$.

To find the equilibrium, we maximise $S(E_m, E_{bh})$ subjected to the constraint that $E_{bh}+E_m=E$ is a constant. That is easily implemented by interpreting $S(E_m,E-E_m)$ as a one-variable function and noting that $\frac{\partial}{\partial E_m}=-\frac{\partial}{\partial E_{bh}}$. The condition $\frac{\partial}{\partial E_m} S(E_m, E-E_m)=0$ is then equivalent to $\frac{\partial S_m}{\partial E_m}=\frac{\partial S_{bh}}{\partial E_{bh}}$, which is, of course, the condition that the temperature of the matter coincides with the temperature of the black hole, in our case, $T=\frac{1}{8\pi E_{bh}}$.

The condition on the second derivative, $\frac{\partial^2}{\partial E_m^2} S(E_m,E-E_m)<0$, becomes
\begin{equation}
    \frac{\partial^2 S_m}{\partial E_m^2}+\frac{\partial^2 S_{bh}}{\partial E_{bh}^2}<0.
    \label{secondd}
\end{equation}

Using $\frac{\partial S}{\partial E}=\frac{1}{T}$ and the formulae for entropy and temperature of a Schwarzschild black hole, eq.~(\ref{secondd}) simplifies to
\begin{equation}
    \frac{1}{E_{bh}}<\frac{1}{T}\frac{\partial T}{\partial E_m},
    \label{seconddsimp}
\end{equation}
or, after using the equation of state~(\ref{eostate}),
\begin{equation}
    E_m<\frac{E_{bh}}{r}.
    \label{maxenergy}
\end{equation}

Equation~(\ref{maxenergy}) establishes the maximum energy the surrounding matter can hold. To translate that condition into entropy, we integrate $T\mathrm{d}S_m=\mathrm{d} E_m$ using eq.~(\ref{eostate}) to obtain $S_m=\frac{r}{r-1}\frac{E_m}{T}$. Hence,
\begin{equation}
\frac{S_m}{S_{bh}}=\frac{2r}{r-1}\frac{E_m}{E_{bh}}<\frac{2}{r-1},
    \label{maxentropy}
\end{equation}
where we used eq.~(\ref{maxenergy}) on the last step.

\begin{figure}[ht]
    \centering
    \includegraphics[width=.6\textwidth]{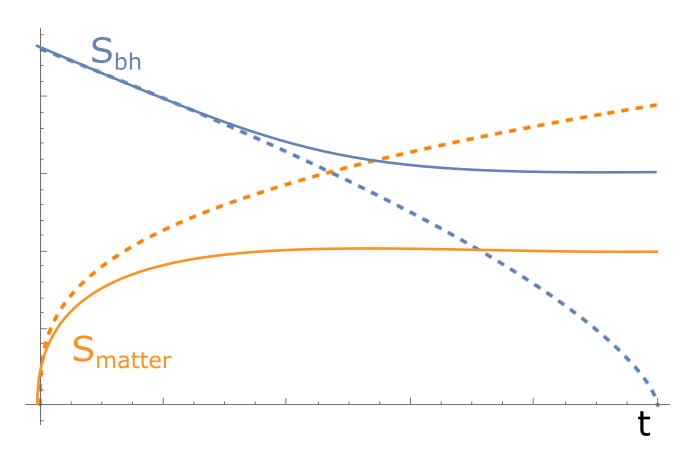}
    \caption{For an unconfined black hole, its entropy (blue dashed line) becomes smaller than the entropy (orange dashed line) of the matter after the Page time. If the system is confined so that it reaches equilibrium with massless particles, the curves representing the black hole entropy (solid blue) and the exterior matter (solid orange) never intercept each other.}
    \label{fig:enter-label}
\end{figure}

Photons are described by the equation of state~(\ref{eostate}) for $\alpha=\frac{\pi^2}{15}$ and $r=4$, which is the Stefan-Boltzmann law. For this value of $r$, eq.~(\ref{maxentropy}) says $\frac{S_m}{S_{bh}}<\frac{2}{3}$, meaning that the Page-time paradox is absent. In other words, the entropy curves never intercept (as sketched in Figure~(\ref{fig:enter-label})). Instead, the system reaches equilibrium before the matter can become more entropic than the black hole.

Indeed, the Page-time paradox can only be realised for black holes that settle down to an equilibrium state with matter confined in a box for equations of state with $r<3$. Such equations of state are implausible. To see why, let us express the condition $r<3$ in terms of the ratio $q$ between the proper pressure $P$ and energy density $u$, $P=qu$. Using the thermodynamical identity
\[\left(\frac{\partial E}{\partial V}\right)_T=T\left(\frac{\partial P}{\partial T}\right)_V-P,\]
we can find $u$ as a function of the temperature assuming $q\neq0$,
\begin{equation}
u\propto T^\frac{1+q}{q},
    \label{gensb}
\end{equation}
which, upon comparing with eq.~(\ref{eostate}), leads to
\begin{equation}
    r=\frac{1+q}{q}.
\end{equation}

Hence, the condition $r<3$ translates into $q>\frac{1}{2}$. A first observation is that the dominant energy condition disallows $q>1$. While violations of this hypothesis are known, especially in quantum field theory, macroscopic breaches would entail faster-than-light energy flows (see, e.g., Ref.~\cite{HE} \textsection 4.3 for a more precise formulation) and are not expected to be physical.

Even though we cannot rule out the interval $\frac{1}{2}<q<1$ on such general grounds, standard kinetic theory arguments (see, e.g. Ref.~\cite{Landau2} \textsection 35) show that if matter behaves approximately in a classical regime governed by particles and electromagnetic fields (other short-ranged interactions are permissible as long as their binding energy is large in magnitude compared to the temperature), one must have $q<\frac{1}{3}$; conversely, the Page-time paradox can only be present for a system that reaches equilibrium when the exterior matter's behaviours deviate considerably from such a classical description.

\section{Conclusion}
In this essay, I briefly point out the strengths and weaknesses of each premise that lead to the Page-tome paradox. I argue that the motivation to hold premise 2 is insufficient. Then, I draw attention to the often underappreciated recognition that an evaporating black hole and its Hawking radiation generally do not constitute a system expected to reach thermodynamic equilibrium. As such, the system evades the applicability of some celebrated results of ordinary and black hole thermodynamics, debilitating premise 4's plausibility. A setting modification that permits the final state to be in equilibrium erases the paradox under reasonable hypotheses, further questioning premise 4.

Hence, I maintain that the Page-time paradox does not require us to revise our understanding of conventional physics, be it gravity, quantum field theory, or statistical mechanics. Still, as often in science, confusion foreshadows learning. For the last half-century, black holes have spearheaded advances in theoretical physics. The confusion around the Page-time paradox might signal the onset of gravitational physics paying back thermodynamics, as it may encourage a systematic treatment of the thermodynamics of entangled subsystems and bring about guidelines for dealing with systems that never reach equilibrium.

\section*{Acknowledgements}
I am thankful for the insightful exchanges with R. Correa da Silva, P. Pessoa and G. Matsas. I am grateful to D. Ryder for pointing out Ref.~\cite{prunkl19} to me, for thoroughly reading a draft of this text and for helping me sharpen its argumentation and improve its style. Other valuable remarks came from D. Sudarsky and Y. Bonder.

\bibliographystyle{unsrt}
\bibliography{references}

\end{document}